\documentclass[a4paper,11pt]{article}

\usepackage{jheppub} 

\usepackage[T1]{fontenc} 
\usepackage{multirow}
\usepackage{mathrsfs}
\usepackage{amsmath}
\usepackage{natbib}
\usepackage{hyperref}
\newcommand{\tabincell}[2]{\begin{tabular}{@{}#1@{}}#2\end{tabular}}

\title{\boldmath Gravity potential determination based on China Space Station Dual-frequency microwave links frequency transfer
}

\author[a,b,c]{Peng Fei Zhang,}
\author[a,c,*]{Chen Xiang Wang,}
\author[a]{Li Hong Li,}
\author[a]{Lei Wang,}
\author[d]{Zi Yu Shen,}
\author[a]{Rui Xu,}
\author[a]{An Ning,}
\author[a,e,f]{Abdelrahim Ruby,}
\author[a,e,*]{and Wen-Bin Shen \note[*]{Corresponding author.}}

\affiliation[a]{School of Geodesy and Geomatics,Wuhan University,\\ Wuhan 430079, China}
\affiliation[b]{Key Laboratory of Surveying and Mapping Science and Geospatial Information Technology of MNR,\\ Beijing 100036, China,}
\affiliation[c]{College of Geographical Sciences, Faculty of Geographical Science and Engineering, Henan University,\\ Zhengzhou, 450046, China}
\affiliation[d]{School of Resource, Environmental Science and Engineering, Hubei University of Science and Technology, \\Xianning, Hubei, China}
\affiliation[e]{State Key Laboratory of Information Engineering in Surveying, Mapping and Remote Sensing, Wuhan University,\\Wuhan 430079, China}
\affiliation[f]{Department of Surveying Engineering, Faculty of Engineering at Shoubra, Benha University,\\Cairo 11629, Egypt}

\emailAdd{pfzhang@sgg.whu.edu.cn}
\emailAdd{wangchenxiang@whu.edu.cn}
\emailAdd{sggllh@whu.edu.cn}
\emailAdd{lwangsgg@whu.edu.cn}
\emailAdd{theorhythm@foxmail.com}
\emailAdd{rXu\_sgg@whu.edu.cn}
\emailAdd{ningan2016@whu.edu.cn}
\emailAdd{abdelrahim.ruby@feng.bu.edu.eg}
\emailAdd{wbshen@sgg.whu.edu.cn}

\abstract{
	The China Space Station (CSS) is currently in orbit and carries the high-precision optical atomic clock with stability of approximately $2.0 \times 10^{-15} / \sqrt{\tau}$ in its experiment module. We have developed a model to determine the gravity potential (GP) based on the gravity frequency shift equation and have created both one-way and dual-frequency transfer models up to $c^{-4}$. These models consider effects from the troposphere, ionosphere and solid Earth tides. The proposed model is suitable for measurements at the magnitude of $10^{-19}$. Based on the CSS mission, we conducted the simulation experiments. The results indicate that when processing the simulation frequency signal using the proposed model, we can obtain the GP with the accuracies of $ (1.13\pm0.71)\,\mathrm{m^2/s^2}$, $ (0.09\pm0.89)\,\mathrm{m^2/s^2}$ and $(0.66\pm1.18)\,\mathrm{m^2/s^2}$ for cutoff elevation angles of $5^{\circ}$, $10^{\circ}$ and $15^{\circ}$ respectively. 
	With the high-precision optical atomic clock onboard the CSS, the proposed model enables us to measure the GP differences in the magnitude of centimeter-level accuracy.
}


\begin{document} 
	\maketitle
	\flushbottom
	
	\section{Introduction}\label{intruduction}
	The Earth has a complicated spatial structure and mass distribution, with its shape evolving primarily under the effect of gravity. Accurately determining the gravity field and the Earth's figure is especially important for various applications. 
	To achieve this, numerous space geodetic techniques have been employed, including very long baseline interferometry (VLBI), Global Navigation Satellite System (GNSS), and Interferometric Synthetic Aperture Radar (InSAR), among some other techniques. In 1983, Will proposed the application of General Relativity Theory (GRT) to gravimetry~\cite{einstein1915,delva2017,Will2018,Mougiakakos2024}. Subsequently, Bjerhammar (1985) introduced the concept of the relativistic geoid, defining it as the surface on which precise clocks run at the same rate, and that is closest to the mean sea level, referred to as the chronometric geoid~\cite{bjerhammar1985}. This definition has two explanations: first, an ideal clock maintains a consistent timescale at an arbitrary point on the chronometric geoid; second, the ideal clock runs with the same oscillation frequency across this chronometric geoid~\cite{shen1993}. 
	This definition offers a novel approach for utilizing GRT to determine gravity potential (GP) differences. Considering the different definitions and types of measurement data, they can be divided into time transfer and frequency transfer, respectively.
	The new methodology for determining GP through time-frequency transfer necessitates atomic clocks with high precision and stability. If the atomic clock accuracy reaches the magnitude of $10^{-18}$, it becomes possible to determine height differences of 1 cm~\cite{delva2017t,denker2018}. However, due to the constraints in the accuracy and stability of high-precision atomic clocks, this method did not receive much attention for an extended period. From the introduction of the first cesium (Cs) atomic clock to define the second in 1967, the relative uncertainties of the best Cs atomic clocks have now approached $10^{-16}$~\cite{guena2017}. In the past decade, optical atomic clocks (OACs) have developed very quickly, achieving significant improvements in long-term stability, now reaching $10^{-19}$~\cite{oelker2019,mcgrew2018, Zhang2023}.
	The development of high-precision atomic clocks has promoted research on the applications of the gravity frequency shift approach for determining the GP. 
	
	In recent decades, researchers have been studying gravitational red-shift (GRS) and chronometric geodesy with the developments of atomic clocks. Several methods have been developed to transfer time and frequency between two points in order to determine the difference in GP. Including, for instance, the Doppler cancellation technique (DCT) for frequency transfer~\cite{vessot1980,shen2016}, clock transportation for time comparison~\cite{Clock2024, Takamoto2022, Grotti2018}, optical fiber/coaxial cables time and frequency comparisons~\cite{Hoang2022, Hoang2021}, precision point position (PPP) time and frequency transfer~\cite{Xu2022,cai2020,defraigne2008}, two-way satellite time and frequency transfer (TWSTFT)~\cite{cheng2022, Huang2016, Jiang2019} and VLBI time transfer~\cite{Wu2021,wang2019}.
	Vessot et al. (1980) adopted the Doppler elimination method to test GRS with a magnitude $7 \times10^{-5}$~\cite{vessot1980}. Bjerhammar proposed the GP determination by using the clock transportation method~\cite{bjerhammar1985}. An alternative approach involves connecting two clocks located at different stations using optical fiber or coaxial cables. Chou et al. (2010) measured the height differences with an accuracy of  $(37\pm 15\, \mathrm{cm})$ by comparing two separate OACs connected via optical fiber~\cite{chou2010}. Results from a transportable optical lattice clock time and frequency transfer experiment operated at Tokyo Skytree indicated that the GRS test achieved an accuracy of $(1.4 \pm 9.1)\times 10^{-5}$~\cite{takamoto2020}. Recently, scholars also determined the gravitational potential by comparing the change in time difference using the PPP time transfer method, whose accuracy reaches $(0.76 \pm 1.79) \, \mathrm{m^2/s^2}$~\cite{cai2020}. 
	The accuracy of height difference measurements using TWSTFT is approximately $(28.0 \pm 5.4) \, \mathrm{m}$~\cite{cheng2022}. 
	Near the surface of Earth, one centimeter height variation could be sensed when the clock comparison accuracy achieves $1\times 10^{-18}$ level ~\cite{mcgrew2018}.
	
	The clock transportation method and the fiber/coaxial cable time and frequency comparison method can only be applied in locations not separated by oceans. The methods of DCT, PPP, and TWSTFT take time and frequency comparison by the satellite, but their applications are constrained by the precision and stability of the satellite clocks. In recent years, many international space atomic clock projects have been promoted one after another to carry high-precision atomic clocks into space. Among these, the Atomic Clock Ensemble in Space (ACES), onboard the International Space Station (ISS) led by the European Space Agency (ESA) will be equipped with atomic clocks offering long-term stability on the magnitude of $10^{-16}$~\cite{cacciapuoti2011,meynadier2018}. The China Space Station (CSS) payloades the OAC with the long-term stability of $3\times10^{-17} @~4000~s$~\cite{guo2021, Guoym2022}. Both missions use microwave links (MWLs) for remote time and frequency comparisons. By implementing these space atomic clock projects, scientists may measure the frequency at a higher accuracy level and develop applications for determining the GP differences.
	
	This paper consists of five sections. Section 1 introduces the relevant background information on chronometric geodesy, atomic clocks, and international space atomic clock projects. In Section 2, we introduce the CSS and its high-precision time and frequency system (HPTFS) carried by CSS. 
	Section 3 provided the relationship between frequency differences and GP differences, presenting models of one-way and dual-frequency transfers to the order of $c^{-4}$, taking into account the effects of the ionosphere, troposphere, and Earth's solid tide. 
	In Section 4, we conduct simulation experiments and analyze the accuracy of GP calculated by the proposed model. Finally, we conclude this work in the last section.
	
	\section{The China Space Station mission}\label{css}
	\subsection{China Space Station }\label{css_1}
\begin{figure*}[htb]
		\centering
		\includegraphics[width=1\textwidth]{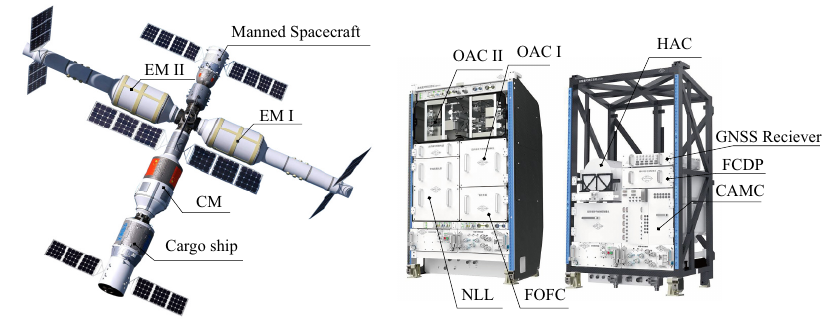}
		\caption{
			The CSS is composed of five parts: core module (CM), experiment module I (EM I), experiment module II (EM II), manned spacecraft and cargo ship. The high-precision time and frequency cabinets are located in the CM II, and they are equipped with a cold atom microwave clock (CAMC), an active hydrogen atomic clock (HAC), and an optical atomic clock (OAC)(Modified after \cite{Css}).}
		\label{fig_CSS}       
	\end{figure*}

	The CSS is designed in a T-shape configuration consisting of five modules: the core module (CM) named Tianhe, experiments module I (EM I) named Wentian, experiments module II (EM II) named Mengtian, a cargo ship and a manned spacecraft~\cite{yang2018} (see in Figure~\ref{fig_CSS}). Tianhe launched on April 29, 2021, serves as the control center of the CSS and has been in orbit for a period of time. Both the CM and EM I are equipped with robotic arms to assist the astronauts during extravehicular activities (EVA)~\cite{yang2018}. The EM I and EM II were launched in 2022 and docked with the CM in the air. Once the experiment modules are assembled, they are used to make the external and internal experiments in the future. The CSS flies at the height of 350$\sim$450 km~\cite{yang2018, Shen2023} with an inclination of $41.5^{\circ}$ orbits. The astronauts and payload devices visit the CSS every six months via manned spacecraft or cargo ships.
	
	The CSS is equipped with a total of thirteen cabinets, which are distributed across different experimental modules. The CM contains two cabinets dedicated to containerless materials and high-microgravity science experiments. The EM I houses four cabinets, primarily used for life science-related experiments. The EM II includes seven cabinets, which are designated for various experiments, including ultra-cold atomic physics, high-precision time and frequency, high-temperature materials, fluid physics, and combustion science.
	
	The HPTFS occupies 1.5 cabinets in the EM II~(see Figure~\ref{fig_CSS}). The system includes an active hydrogen atomic clock (HAC), a cold atomic microwave clock (CAMC), and an OAC. The OAC is composed of the electronic part (OAC I in Figure~\ref{fig_CSS}) and the physical part (OAC II in Figure~\ref{fig_CSS}). The performances of the three clocks are shown in table~\ref{stability}. 
	The active HAC outputs frequency and second pulse signals with medium and long-term stability, while the CMAC and OAC provide frequency signals with medium and long-term stability. These clocks provide frequency references for relevant loads. The HPTFS generates high-precision time and frequency signals with second stability better than $2\times10^{-15}$ (See in Table~\ref{stability}), daily stability in the magnitude of $10^{-17}$~\cite{guo2021,sun2021, Guoym2022}.  Additionally, the experimental cabinets support both microwave and laser time-frequency transfer. The stability is approximately 0.3 ps at 300s and 6 ps at one day for microwave time-frequency transfer, while the laser time-frequency transfer has a stability of about 0.1 ps at 300 s and 1 ps at one day.  In this study, we will focus on using the microwave time-frequency transfer method to determine the GP.
	
	The onboard comparison between the OAC and CAMC is ensured by the Frequency Comparison and Distribution Package (FCDP)~\cite{cacciapuoti2009, Shen2023}. 
	A GNSS receiver is installed in the HPTFS and connected to the CMAC, providing precise orbital data for HPTFS.
	The Femtosecond Optical Frequency Comb (FOFC)~\cite{katori2011,zhu2019,wei2009} is the core component of the HPTFS. It performs several key tasks, including the downconversion of the space cold-atom optical clock frequency, referencing and measuring the laser frequency of the space OAC, supplying ultra-stable microwave signals to the CAMC, and completing the frequency upconversion of both the CAMC and HAC, etc. Combined with the FOFC and Narrow line laser (NLL), the OAC can provide high-precision time-frequency signals. 
	
	\begin{table}[htb]
		\centering
		\begin{tabular}{|l|c|c|}
			\hline
			Clocks and Links&Short-term 
			Stability &{Long-term Stability} \\
			\hline
			Hydrogen Atomic Clock&$2\times10^{-13}@ 1s$&$2\times10^{-15}@ 1day$\cr
			Cold Atomic Microwave Clock&$5\times10^{-14}@ 1s$&$2\times10^{-16}@ 1day$\cr
			Optical Atomic Clock&$2\times10^{-15}@ 1s$&$3\times10^{-17}@ 4000s$\cr
			Microwave time-frequency transfer&$0.3ps@300s$ &$6ps@ 1day$\cr
			Laser time-frequency transfer&$0.1ps@300s$ &$1ps@ 1day$\cr
			\hline
		\end{tabular}
		\caption{Main technical indicators of the high-precision time and frequency experiment cabinet.  Data come from the China Space Station science experiment resource manual (\url{www.csu.cas.cn/gb/201905/P020190507639578655422.pdf}). }
		\label{stability}       
	\end{table}
	
	The CSS is expected to remain in orbits for about ten years, with plans to potentially replace its current clock in the HPTFS with a ytterbium ($\mathrm{Yb}$) OAC whose long-term stability is about $10^{-19}$~\cite{lv2021}. High-precision orbit determination unit can achieve an orbital accuracy of better than 10  cm in radial direction and an accuracy of velocity better than 1 mm/s after post-processing. These will promote the development of GRT and relativistic geodesy. Using the current OAC on board the CSS whose long-term stability is $3\times10^{-17}$, we may measure the GP at 3 dm level. Once the CSS is equipped with an OAC at the $10^{-19}$ level, the GP determination of the ground station (GS) could be determined at 1 cm level.
	
	\subsection{Microwave links} \label{css_2}
	
	The MWLs can be used for time and frequency comparison between the CSS and GS. It consists of a flight segment in the CSS and several ground segments. Currently, there are three national GSs (Beijing, Xi'an, and Shanghai), while a fourth national GS is being planned in Wuhan. All these stations are equipped with MWL devices.
	
	The MWL system has four work frequencies which are distributed by FCDP. It consists of two uplinks, both transmitting left-hand circularly polarized (LHCP) signals at frequencies $26.8$ GHz and $30.4$ GHz, and two downlinks that transmit right-hand circularly polarized (RHCP) signals at frequencies $20.8$ GHz and $30.4$ GHz, respectively. Notably, the MWL signals with frequency $30.4$ GHz have different polarization directions for uplink and downlink (see Table~\ref{tab_mwl}). The stability of MWL is determined by the clocks of CSS and sent by a circularly polarized antenna which is used for both transmission and reception. The microwave beam width of the antenna is $\pm 70^{\circ}$. Furthermore, the maximum phase center variation is controlled within $5^{\circ}$, ensuring that the phase center stability remains below 0.6 mm. By comparing the HPTFS reference frequency to a set of ground clocks, it becomes possible to conduct tests on GRS~\cite{sun2021,cacciapuoti2011}, measure GP differences~\cite{cai2020,tanaka2021} and explore other applications.
	
	\begin{table}[!h]
		\centering
		\begin{tabular}{|l|l l|}
			\hline
			Parameters & \multicolumn{2}{|c|}{Specifications}  \\
			\hline
			\multirow{4}{*}{Work frequencies}&\multirow{2}{*}{Recieved}& \tabincell{c}{$30.
				4\,\mathrm{GHz}$ (LHCP)}\cr
			&\multirow{2}{*}{}& \tabincell{c}{$26.8\,\mathrm{GHz}$ (LHCP)}\cr
			&\multirow{2}{*}{Emitted}& \tabincell{c}{$30.4\,\mathrm{GHz}$ (RHCP)}\cr
			&\multirow{2}{*}{}& \tabincell{c}{$20.8\, \mathrm{GHz} $ (RHCP)}\cr
			{Phase center stability}&\multicolumn{2}{|c|}{$\leq 0.6 \thinspace mm\,(RMS)$}\cr
			\hline
		\end{tabular}
		\caption{MWL's different design technical specifications}
		\label{tab_mwl}       
	\end{table}
	
	\section{Method}\label{method}
	Compared with the traditional method of determining GP which needs combining the geometric leveling and gravity~\cite{meynadier2018},
	This study introduced an alternative approach to measure the GP by gravity frequency shift (GFS)~\cite{delva2017}, by which we may determine the GP directly. In the following sections, we will introduce the fundamental theory of determining the GP by MWL frequency transfer.
	
	\subsection{One-Way frequency transfer to order $c^{-4}$}\label{method_2}
	When the stability of the clock reaches $10^{-18}$, according to the Post-Newtonian approximation of the metric theories of gravity, the accuracy of one-way frequency transfer must be at the order of $c^{-4}$~\cite{linet2002,Wu2023gra}. In this section, we will create a one-way frequency transfer model in the order of $c^{-4}$, which will contain the mass and spin multipole moments of the isolated, axisymmetric rotating body, ionospheric and tropospheric effects.
	\begin{figure}[htb]
		\centering
		\includegraphics[width=1\textwidth]{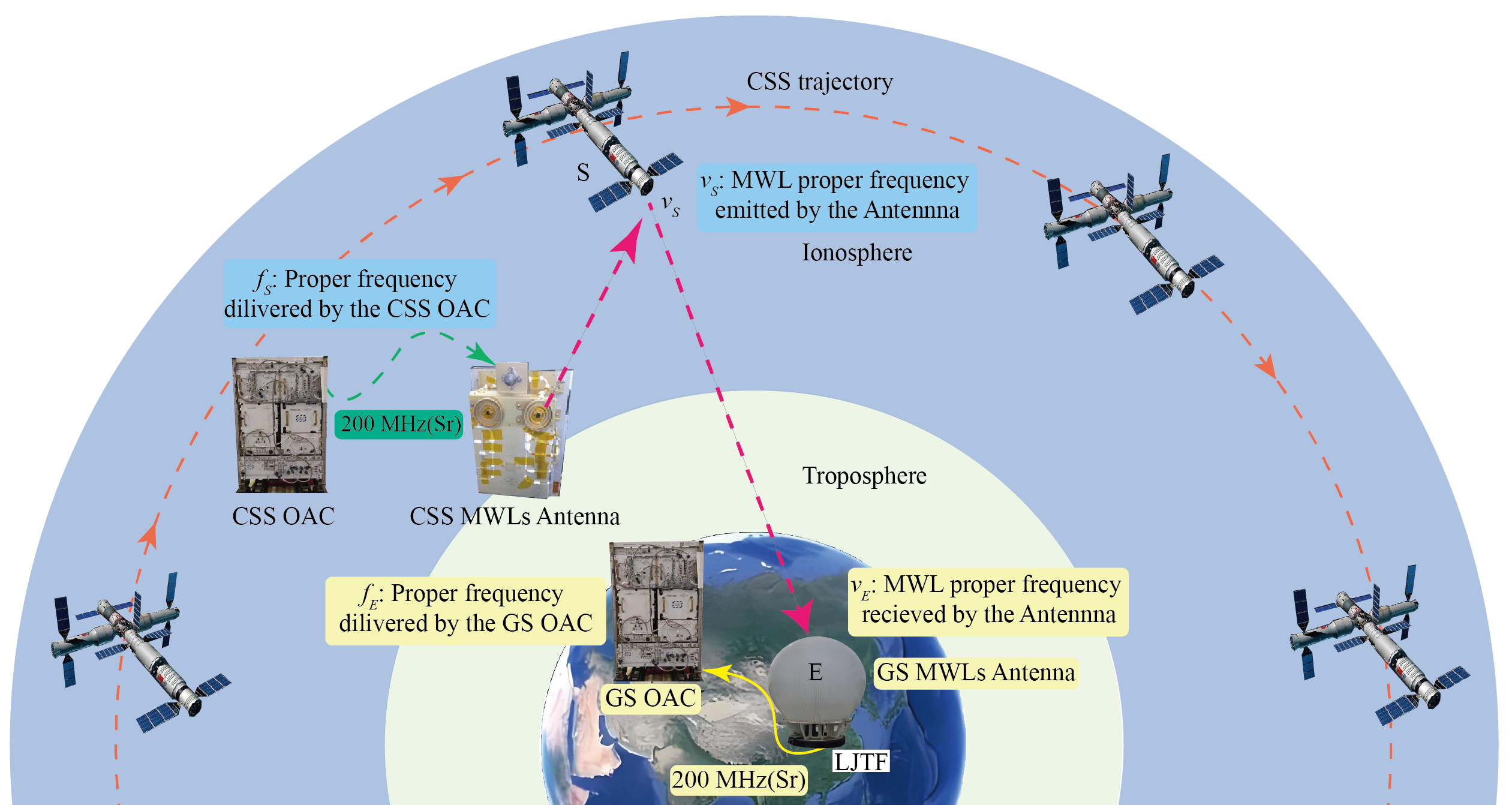}
		\caption{One-way frequency transfer for the MWL. The OAC of the CSS generates a signal with the proper frequency $f_{S}$, and the FOFC then performs downconversion, outputting a 200 MHz signal to the antenna, which subsequently emits the MWL signal with frequency $\nu_S$ to the GS. At the CSS, we can obtain the ratio $f_{S}/\nu_{S}$ or $\nu_{S}/f_{S}$. When the MWL reaches the GS antenna, the signal frequency becomes $\nu_E$. Through upconversion by the FOFC, this signal can be compared with the OAC frequency of the GS, allowing us to get the values of $f_{E}/\nu_{E}$ or $\nu_{E}/f_{E}$.}
		\label{fig_one}
	\end{figure}
	
	This discussion considers a one-way frequency signal link between the CSS and GS. The main task of frequency transfer is determining the ratio of $f_{S}$ and $f_{E}$ between the clocks located at points S and E, where $f_{S}$ and $f_{E}$ represent the proper frequencies delivered by the clock (S) and clock (E), respectively.
	When photons are sent from clock $S$ to clock $E$, the proper frequencies of the photons are denoted as $\nu_{S}$ and $\nu_{E}$. The ratio can be expressed as $f_{E} / f_{S}=\left(f_{E} / \nu_{E}\right)\left(\nu_{E} / \nu_{S}\right)\left(\nu_{S} / f_{S}\right)$, where $\nu_{S} / f_{S}$ and $f_{E}/\nu_{E}$ can be measured by the local clocks at S and E~\cite{blanchet2001}. Assuming the signal is emitted at time $t_S$ and received at time $t_E$, in a geocentric inertial (non-rotating) coordinate frame, the positions of S and E can be represented as $\boldsymbol{r}_{\mathrm{S}}=\boldsymbol{x}_{\mathrm{S}}\left(t_{\mathrm{S}}\right)$ and $\boldsymbol{r}_{\mathrm{E}}=\boldsymbol{x}_{\mathrm{E}}\left(t_{\mathrm{E}}\right)$, respectively. It is necessary to take into account terms of order $1/c^{4}$ into the ratio
	of $\nu_{B}/\nu_{A}$ when the frequency shift stability reaches  $10^{-18}$~\cite{linet2002}:
	\begin{align}
		\frac{\delta \nu}{\nu} \equiv \frac{\nu_{E}}{\nu_{S}}-1=\left(\frac{\delta \nu}{\nu}\right)_{c}+\left(\frac{\delta \nu}{\nu}\right)_{g}+O(c^{-5}) \label{1}
	\end{align}
	where $(\delta\nu/\nu)_{c}$ is the special-relativistic Doppler effect, and $(\delta\nu/\nu)_{g}$ contains all the contributions from the gravitational field, which is also mixed with kinetic terms. These effects can expressed in the geocentric inertial coordinate frame as follows:
	\begin{align}
		\left(\frac{\delta \nu}{\nu}\right)_{c}=\sum_{j=1}^{4} \Delta_{c}^{(j)}\nonumber\\
		\left(\frac{\delta \nu}{\nu}\right)_{g}=\sum_{j=2}^{4} \Delta_{g}^{(j)}\label{c2}
	\end{align}
	where the terms $\Delta_{c}$ in equation~(\ref{c2}) are defined~\cite{linet2002}:
	\begin{align}
		\begin{aligned}
			\Delta_{c}^{(1)} =&-\frac{1}{c} \mathbf{N}_{SE} \cdot\left(\boldsymbol{v}_{S}-\boldsymbol{v}_{E}\right) \\
			\Delta_{c}^{(2)} =&-\frac{1}{c^{2}}\left[\frac{1}{2} v_{S}^{2}-\frac{1}{2} v_{E}^{2}+\left[\mathbf{N}_{SE} \cdot\left(\boldsymbol{v}_{S}-\boldsymbol{v}_{E}\right)\right]\left(\mathbf{N}_{SE} \cdot \boldsymbol{v}_{S}\right)\right] \\
			\Delta_{c}^{(3)} =&\frac{1}{c^{3}}\left[\mathbf{N}_{SE} \cdot\left(\boldsymbol{v}_{S}-\boldsymbol{v}_{E}\right)\right]\left[\frac{1}{2}v_{S}^{2}-\frac{1}{2}v_{E}^{2}-\left(\mathbf{N}_{SE} \cdot \boldsymbol{v}_{S}\right)^{2}\right] \\
			\Delta_{c}^{(4)}=&\frac{1}{c^{4}}\left[\frac{3}{8} v_{E}^{4}-\frac{1}{4} v_{S}^{2} v_{E}^{2}-\frac{1}{8} v_{S}^{4}-\left[\mathbf{N}_{SE} \cdot\left(\boldsymbol{v}_{S}-\boldsymbol{v}_{E}\right)\right]\right.\\
			&\left.\left(\mathbf{N}_{SE} \cdot \boldsymbol{v}_{S}\right)\left(\frac{1}{2} v_{S}^{2}-\frac{1}{2} v_{E}^{2}-\left(\mathbf{N}_{SE} \cdot \boldsymbol{v}_{E}\right)^{2}\right)\right] 
		\end{aligned}\label{dop_c}
	\end{align}
	where $\mathbf{N}_{\mathrm{SE}}=\left(\mathbf{r}_{\mathrm{E}}-\mathbf{r}_{\mathrm{S}}\right) /\left|r_{\mathrm{E}}-r_{\mathrm{S}}\right|$; $\boldsymbol{v}_S$ and $\boldsymbol{v}_E$ are the velocity vectors of the CSS and GS, respectively,  and their absolute values are $v_S$ and $v_E$. 
	In the CSS mission, we find the numerical contributions for four terms caused by position and velocity. The first term, often called as the first-order Doppler effect~\cite{vessot1980}, is about $\left|\Delta_{c}^{(1)}\right|<2.23 \times 10^{-5}$. If the positions of the CSS and GS remain unchanged, the first-order Doppler frequency shifts for both uplink and downlink are equal, which can be eliminated through the differential of uplink and downlink signals. The second term contains the second-order Doppler effect, with a value range of approximately $\left|\Delta_{c}^{(2)}\right|<3.56 \times 10^{-10}$. The third and fourth terms  numerical contribution are about $\left|\Delta_{c}^{(3)}\right|<7.33 \times 10^{-15}$ and $\left|\Delta_{c}^{(4)}\right|<1.64 \times 10^{-19}$, respectivily.
	
	When we ignored items with magnitudes less than $10^{-19}$ in the $\Delta_{g}^{(4)}$ term~\cite{linet2002}, the terms $\Delta_{g}$ of equation~(\ref{c2}) can be expressed:
	\begin{align}
		\begin{aligned}
			\Delta_{g}^{(2)} =&-\frac{1}{c^{2}}\left(U_{S}-U_{E}\right) \\
			\Delta_{g}^{(3)} =&\frac{1}{c^{3}}\left\{\left(U_{S}-U_{E}\right)\left[\mathbf{N}_{SE} \cdot\left(\boldsymbol{v}_{S}-\boldsymbol{v}_{E}\right)\right]-\boldsymbol{l}_{S}\cdot \boldsymbol{v}_{S}+\boldsymbol{l}_{E} \cdot \boldsymbol{v}_{E}\right\} \\
			\Delta_{g}^{(4)} =&-\frac{1}{c^{4}}\left.\{(\gamma+1)\left(U_{S} v_{S}^{2}-U_{E} v_{E}^{2}\right)-\frac{1}{2}\left(U_{S}-U_{E}\right)\right.\\
			& \times\left\{U_{S}-U_{B}-2(1-\beta)\left(U_{S}+U_{E}\right)+v_{S}^{2}-v_{E}^{2}\right.\\
			&\left.\left.-2\left[\mathbf{N}_{SE} \cdot\left(\boldsymbol{v}_{S}-\boldsymbol{v}_{E}\right)\right]\left(\mathbf{N}_{SE} \cdot \boldsymbol{v}_{S}\right)\right\}\right\}
		\end{aligned}\label{dop_g}
	\end{align}
	where $U_S$ and $U_E$ are the gravitational potentials of points S and E, respectively, there are two nonvanishing post$-$Newtonian parameters, $\gamma=1$ and $\beta=0$. This indicates that equation (\ref{dop_g}) applies solely to stationary gravitational fields~\cite{linet2002}. $\boldsymbol{l}_{S}$ and $\boldsymbol{l}_{E}$ are given up to the order defined by the equations:
	\begin{align}
		\boldsymbol{l}_{S} =c^{2}\times[\boldsymbol{l}_{M}\left(\boldsymbol{x}_{S}, \boldsymbol{x}_{E}\right)+\boldsymbol{l}_{J_{2}}\left(\boldsymbol{x}_{S}, \boldsymbol{x}_{E}\right)+\cdots]\label{la}
	\end{align}
	\begin{align}
		\boldsymbol{l}_{E} =c^{2}\times[-\boldsymbol{l}_{M}\left(\boldsymbol{x}_{E}, \boldsymbol{x}_{S}\right)-\boldsymbol{l}_{J_{2}}\left(\boldsymbol{x}_{E}, \boldsymbol{x}_{S}\right)+\cdots] \label{lb}
	\end{align}
	where $\boldsymbol{l}_{M}$ and $\boldsymbol{l}_{J_2}$ represent the mass and quadrupole moment yield contributions to the order $1/c^{-2}$, the expression of them can be found in Linet's study~\cite{,linet2002}. We evaluate the influence of these three terms in equation~(\ref{dop_g}), specifically:
	$\left|\Delta_{g}^{(2)}\right|<3.90 \times 10^{-11}$,
	$\left|\Delta_{g}^{(3)}\right|<3.12 \times 10^{-14}$, $\left|\Delta_{g}^{(4)}\right|<5.08 \times 10^{-19}$.
	
	According to equations~(\ref{1}) to~(\ref{lb}), we can calculate the frequency shift to the order of $10^{-18}$ in the vacuum. However, in the space of Earth, MWLs are influenced by the atmosphere (including troposphere and ionosphere), Earth and celestial tides. The tidal effects of the CSS and GS are contained in $U_S$ and $U_E$; their numerical values are less than $3.34\, m^2/s^2$ and $5.71\thinspace m^2/s^2$, respectively. It is essential to eliminate or correct these influences in our methodologies. The previous studies indicate that when an MWL transfers from the CSS to GS, the frequency shift ($\Delta f$) is caused by the time variation of the phase path~\cite{jacobs1966,bennett1968}:
	\begin{equation}
		\Delta f=-\frac{f}{c} \frac{d P}{d t} \label{df}
	\end{equation}
	where $c$ denotes the speed of light in the vacuum, $f$ represents the proper frequency of the MWL, and  $P$ signifies the phase path. We use $n_i$ and $n_t$ to refer to the refractive indices of the ionosphere and troposphere, respectively, while $L_i$ and $L_t$ are the propagation path of the MWL through the ionosphere and troposphere. Consequently, the frequency shifts caused by the atmosphere can be expressed as~\cite{millman1984,shen2016,shen2021f}:
	\begin{equation}
		\Delta f=\Delta f_{\text {ion}}+\Delta f_{\text {tro}}=-\frac{f}{c}\frac{d}{dt}\int_{L_i}(n_{i}-1)dl_i-\frac{f}{c}\frac{d}{dt}\int_{L_t}(n_{t}-1)dl_t \label{dfn}
	\end{equation}
	
	The ionospheric refractive index $n_i$ can be expressed as~\cite{hoque2007,hoque2012}:
	\begin{equation}
		\begin{aligned}
			n_i=&1-40.3 \frac{N_{e}}{f^{2}} \pm \frac{7527 \times c}{2 f^{3}} N_{e} B_{0} \cos \theta -\frac{812.3}{f^{4}} N_{e}^{2}-\frac{1.58 \times 10^{22}}{f^{4}} N_{e} B_{0}^{2}\left(1+\cos ^{2} \theta\right)
		\end{aligned}\label{ni}
	\end{equation}
	where 
	$N_e$ represents the ionospheric electron density, $B_0$ denotes the strength of the geomagnetic field, $\theta$ is the angle between the direction of wave propagation and the geomagnetic field vector. When the MWL transmits a LHCP signal, the sign is positive ($+$); conversely, if the MWL is a
	RHCP signal, the sign is negative ($-$). Since the third-order ionospheric frequency shifts are less than $10^{-20}$~\cite{Zhangpf2023}, we only consider the second-order ionospheric frequency shifts.  By substituting equation (\ref{ni}) into the first term on the left side of equation~(\ref{dfn}) and only considering the $f^{-2}$ and $f^{-3}$ terms, we obtain~\cite{Zhangpf2023}:
	\begin{equation}
		\Delta f_{\text {ion}}=40.3 \frac{1}{c f} \frac{\mathrm{d}}{\mathrm{d} t} \int_{L_{i}} N_{e} d l_{i} \mp \frac{7527}{2 f^{2}} \frac{\mathrm{d}}{\mathrm{d} t} \int_{L_{i}} N_{e} B_{0} \cos \theta d l_{i} \label{dfion}
	\end{equation}
	
	The tropospheric refractive index $n_t$ is described as~\cite{rueger2002,shen2021f}: 
	\begin{equation}
		n_{t}=1+\left(k_{1} \frac{p_{d}}{T}+k_{2} \frac{p_{w}}{T}+k_{3} \frac{p_{w}}{T^{2}}\right) \times 10^{-6}+\epsilon \label{nt}
	\end{equation}
	where $k_1, k_2$ and $k_3$ are constant coefficients, $T$ represents the temperature in Kelvin (K), $p$ denotes the pressure, the subscript $d$ and $w$ mean the dry and wet partial pressures, respectively. Additionally, $\epsilon$ signifies the uncertainty. By taking the $n_t$ into the second term on the left side of equation~(\ref{dfn}), we can derive the tropospheric frequency shift:
	\begin{equation}
		\Delta f_{\text {tro}}=-\frac{f}{c} \frac{\mathrm{d}}{\mathrm{d} t} \int_{L_{t}}\left(\left(k_{1} \frac{p_{d}}{T}+k_{2} \frac{p_{w}}{T}+k_{3} \frac{p_{w}}{T^{2}}\right) \times 10^{-6}+\epsilon
		\right) d l_{t} \label{dftro}
	\end{equation}
	
	When considering the MWL frequency $f$, we can obtain the relative tropospheric and ionospheric frequency shifts as $\Delta f_{\text {tro}}/f$ and $f_{\text {ion}}/f$, respectively.
	
	\subsection{Dual-frequency transfer for the CSS mission}\label{method_3}
	For one-way frequency transfer, the ionospheric and tropospheric frequency shifts are represented as $\Delta f_{ion}/f$ and $\Delta f_{tro}/f$, respectively. Combining equations~(\ref{dfion}) and~(\ref{dftro}), we can see that the ionospheric frequency shifts are affected by the frequency of signals, while the tropospheric frequency shifts do not affect signal frequency. To eliminate or correct the frequency shifts along the propagation path, an X combination of uplink and downlink is employed in frequency transfer. 
	\begin{figure}[htb]
		\centering
		\includegraphics[width=1\textwidth]{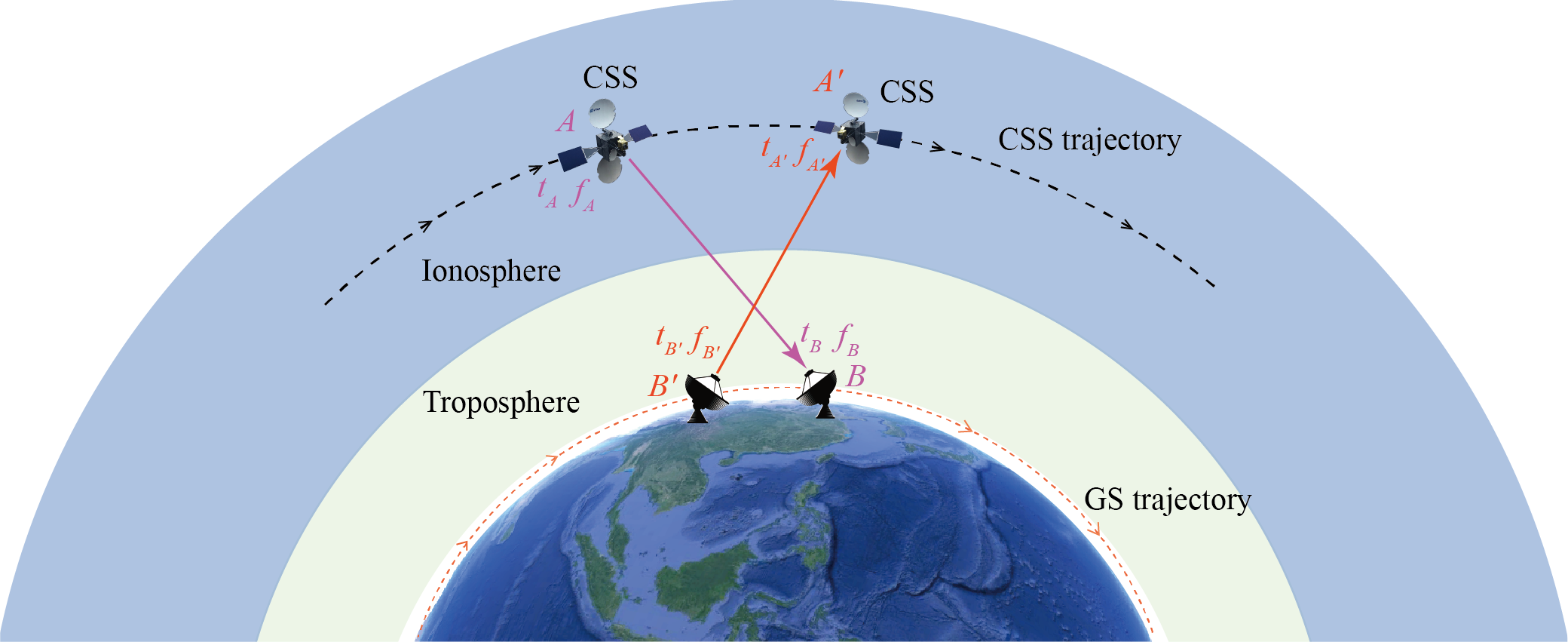}
		\caption{Dual-frequency time and frequency transfer for the CSS mission. The GS emits a MWL with frequency $f_{B^{'}}$ at time $t_{B^{'}}$, which isreceived by the CSS as the signal $f_{A^{'}}$ at $t_{A^{'}}$. Simultaneously, the CSS transmits a MWL with frequency $f_{A}$ at time $t_{A}$ and the GS receives the signal $f_{B}$ at time $t_{B}$.}
		\label{fig_MWL}
	\end{figure}
	
	The X configuration is briefly illustrated in Figure~\ref{fig_MWL}. In this setup, one MWL is emitted from the GS $B^{\prime}$ at time $t_{B^\prime}$ with frequency $f_{B^\prime}$ and it is received at point $A^\prime$ at time $t_{A^\prime}$ with frequency $f_{A^\prime}$. At the same time, the CSS emits an MWL with frequency $f_{A}$ at time $t_{A}$,  which is received by the GS as frequency $f_{B}$ at time $t_{B}$ in the GS. The time interval between emission times in the CSS and GS is approximately
	$\Delta t=t_{A}-t_{B^{'}}\leq 1 \times 10^{-6}\space s$~\cite{duchayne2009,sun2021}. 
	
	From the study of the section~\ref{method_2}, considering the propagation frequency shifts and tidal effects on the one-way frequency transfer, the expression for the frequency transfer of the uplink is:
	\begin{equation}
		\begin{aligned}
			&\frac{\Delta f_{B^{\prime} A^{\prime}}}{f_{B^{\prime}}}  \equiv \frac{f_{A^{\prime}}}{f_{B^{\prime}}}-1 \\
			&=\left(\frac{\Delta f_{B^{\prime} A^{\prime}}}{f_{B^{\prime}}}\right)_{c}+\left(\frac{\Delta f_{B^{\prime} A^{\prime}}}{f_{B^{\prime}}}\right)_{g}^{(4)}+\frac{\Delta f^{\prime}_{\text {ion}}}{f_{B^{\prime}}}+\frac{\Delta f^{\prime}_{\text {tro}}}{f_{B^{\prime}}}+\frac{1}{c^{2}}\left(U_{A^{\prime}}-U_{B^{\prime}}+\Delta U_{tide_{A^{\prime}B^{\prime}}}\right) \\
			&-\frac{1}{c^{3}}\left\{\left(U_{A^{\prime}}-U_{B^{\prime}}+\Delta U_{tide_{A^{\prime}B^{\prime}}}\right)\left[\mathbf{N}_{A^{\prime} B^{\prime}} \cdot\left(\boldsymbol{v}_{A^{\prime}}-\boldsymbol{v}_{B^{\prime}}\right)\right]-\boldsymbol{l}_{A^{\prime}}^{(2)} \cdot \boldsymbol{v}_{A^{\prime}}+\boldsymbol{l}_{B^{\prime}}^{(2)} \cdot \boldsymbol{v}_{B^{\prime}}\right\}
		\end{aligned} \label{dfba}
	\end{equation}
	and the expression for the downlink MWL is:
	\begin{equation}
		\begin{aligned}
			&\frac{\Delta f_{A B}}{f_{A}} \equiv \frac{f_{B}}{f_{A}}-1 \\
			&=\left(\frac{\Delta f_{A B}}{f_{A}}\right)_{c}+\left(\frac{\Delta f_{AB}}{f_A}\right)_{g}^{(4)}+\frac{\Delta f_{\text {ion }}}{f_A}+\frac{\Delta f_{\text {tro }}}{f_A}-\frac{1}{c^{2}}\left(U_{A}-U_{B}+\Delta U_{tide_{AB}}\right) \\
			&+\frac{1}{c^{3}}\left\{\left(U_{A}-U_{B}+\Delta U_{tide_{AB}}\right)\left[\mathbf{N}_{A B} \cdot\left(\boldsymbol{v}_{A}-\boldsymbol{v}_{B}\right)\right]-\boldsymbol{l}_{A}^{(2)} \cdot \boldsymbol{v}_{A}+\boldsymbol{l}_{B}^{(2)} \cdot \boldsymbol{v}_{B}\right\}
		\end{aligned}\label{dfab}
	\end{equation}
	
	We know that $\left|\Delta_{g}^{(4)}\right|<5.08 \times 10^{-19}$. The gravitational potentials of the GS are calculated by equations~(\ref{dfba}$\sim$\ref{dfab}). We consider $\Delta U_{AB}$ to the order of magnitude $c^{-3}$, and the gravitational potentials of the GS for both uplink and downlink are:   
	\begin{equation}
		\begin{aligned}
			&U_{B^{\prime}}= U_{A^{\prime}}+\Delta U_{\text{tide}_{A^{\prime} B^{\prime}}} \\
			&- \frac{\frac{\Delta f_{B^{\prime} A^{\prime}}}{f_{B^{\prime}}}-\left(\frac{\Delta f_{B^{\prime} A^{\prime}}}{f_{B^{\prime}}}\right)_{c}-\frac{1}{c^{3}}\left\{\boldsymbol{l}_{A}^{(2)} \cdot \boldsymbol{v}_{A}-\boldsymbol{l}_{B^{\prime}}^{(2)} \cdot \boldsymbol{v}_{B^{\prime}}\right\}-\left(\frac{\Delta f_{B^{\prime} A^{\prime}}}{f_{B^{\prime}}}\right)_{g}^{(4)}-\frac{\Delta f^{\prime}_{\text {ion}}}{f_{B}^{\prime}}-\frac{\Delta f^{\prime}_{\text {tro}}}{f_{B}^{\prime}}}{\left[\frac{1}{c^{2}}-\frac{1}{c^{3}} \mathbf{N}_{A^{\prime} B^{\prime}} \cdot\left(\boldsymbol{v}_{A}-\boldsymbol{v}_{B^{\prime}}\right)\right]}
		\end{aligned} \label{ubp}
	\end{equation}
	and
	\begin{equation}
		\begin{aligned}
			&U_{B}= U_{A}+\Delta U_{\text {tide}_{A B}} \\
			+& \frac{\frac{\Delta f_{A B}}{f_{A}}-\left(\frac{\Delta f_{A B}}{f_{A}}\right)_{c}+\frac{1}{c^{3}}\left\{\boldsymbol{l}_{A}^{(2)} \cdot \boldsymbol{v}_{A}-\boldsymbol{l}_{B}^{(2)} \cdot \boldsymbol{v}_{B}\right\}-\left(\frac{\Delta f_{A B}}{f_{A}}\right)_{g}^{(4)}-\frac{\Delta f_{\text {ion}}}{f_{A}}-\frac{\Delta f_{\text {tro}}}{f_{A}}}{\left[\frac{1}{c^{2}}-\frac{1}{c^{3}} \mathbf{N}_{A B} \cdot\left(\boldsymbol{v}_{A}-\boldsymbol{v}_{B}\right)\right]}
		\end{aligned} \label{ub}
	\end{equation}
	
	We evaluate the frequency shifts due to the ionosphere, troposphere, Doppler effect, and the tidal influence on both the CSS and GS. 
	The magnitude of the Doppler frequency shift reaches $10^{-5} \sim 10^{-6}$, while the magnitudes of the ionospheric and tropospheric frequency shifts are approximately $10^{-10} \sim 10^{-13}$ and $10^{-13} \sim 10^{-15}$, respectively. The tidal effects on the CSS and GS are less than
	$1.70 \, m^2/s^2$ and $3.08 \, m^2/s^2$, respectively. From equation (\ref{dfion}), we can observe that ionospheric frequency shifts are related to the polarization directions and frequencies of MWLs. Based on the characteristics of the MWL used in the CSS mission (see Table~\ref{tab_mwl}), we selected the Ka-band signals with the same frequency $30.4$ GHz but with different polarization directions as the working frequency points to make frequency transfer. In the X configuration, the angles $\theta$ in second-order terms of equation~(\ref{dfion}) for uplink and downlink are denoted as $\theta_{u}$ and $\theta_{d}$, respectivily, where $\theta_{u}+\theta_{d} \approx 180^{\circ}$. $\cos \theta_{u}\approx-\cos \theta_{d}$. By combining $\cos \theta_{u}$ and $\cos \theta_{d}$ with their different signs of second-order terms, we find that the ionospheric frequency shifts have similar values.
	
	We focus on the static part of the GP ($W$), which consists of gravitational potential $U_E$ and the centrifugal potential $Z_E$. $W$ can be expressed as~\cite{Mehlstubler2018,muller2018}:
	\begin{equation}
		W=U_{\mathrm{E}}+Z_{\mathrm{E}} \label{W}
	\end{equation}
	the centrifugal potential $Z_{\mathrm{E}}$ in the left side of equation~(\ref{W}) is defined as $Z_{\mathrm{E}}=v^{2}/2$. By combining equations~(\ref{ubp}) to ~(\ref{W}), the GP of GS can be expressed as:
	\begin{equation}
		\begin{aligned}
			&W_{B} \approx \frac{\left(U_{B}+\frac{1}{2} v_B^{2}+U_{B^{\prime}}+\frac{1}{2} v_{B^\prime}^{2}\right)}{2}\\
			&=\frac{1}{2}\left(U_{A}+U_{A^{\prime}}+\delta U_{\text {tide }}+\delta f_{\text {ion }}+\delta f_{\text {tro }}+\frac{1}{2} v_{B^\prime}^{2}+\frac{1}{2} v_{B}^{2}\right)\\
			&+\frac{1}{2} \frac{\frac{\Delta f_{A B}}{f_{A}}-\left(\frac{\Delta f_{A B}}{f_{A}}\right)_{c}-\frac{1}{c^{3}}\left\{\boldsymbol{l}_{B}^{(2)} \cdot \boldsymbol{v}_{B}-\boldsymbol{l}_{A}^{(2)} \cdot \boldsymbol{v}_{A}\right\}-\left(\frac{\Delta f_{A B}}{f_{A}}\right)_{g}^{(4)}}{\left[\frac{1}{c^{2}}+\frac{1}{c^{3}} \mathbf{N}_{AB} \cdot\left(\boldsymbol{v}_{B}-\boldsymbol{v}_{A}\right)\right]}\\
			&-\frac{1}{2} \frac{\frac{\Delta f_{B^{\prime} A^{\prime}}}{f_{B^{\prime}}}-\left(\frac{\Delta f_{B^{\prime} A^{\prime}}}{f_{B^{\prime}}}\right)_{c}-\frac{1}{c^{3}}\left\{\boldsymbol{l}_{A^{\prime}}^{(2)} \cdot \boldsymbol{v}_{A}-\boldsymbol{l}_{B^{\prime}}^{(2)} \cdot \boldsymbol{v}_{B^{\prime}}\right\}-\left(\frac{\Delta f_{B^{\prime} A^{\prime}}}{f_{B^{\prime}}}\right)_{g}^{(4)}}{\left[\frac{1}{c^{2}}-\frac{1}{c^{3}} \mathbf{N}_{A^{\prime} B^{\prime}} \cdot\left(\boldsymbol{v}_{A^{\prime}}-\boldsymbol{v}_{B^{\prime}}\right)\right]}
		\end{aligned} \label{WB}
	\end{equation}
	where $\delta U_{tide}$, $\delta f_{\text {ion }}$ and $\delta f_{\text {tro }}$ represent the residual errors of the tide, ionospheric and tropospheric frequency shifts.
	$\delta U_{t i d e}=\Delta U_{t i d e_{AB}}+\Delta U_{t i d e_{A^{\prime} B^{\prime}}}$, $\delta f_{\text {ion }}=\frac{\Delta f_{\text {ion }}^{\prime}}{f_{B^{\prime}}}-\frac{\Delta f_{\text {ion }}}{f_{A}}$, $\delta f_{\text {tro }}=\frac{\Delta f_{\text {tro }}^{\prime}}{f_{B^{\prime}}}-\frac{\Delta f_{\text {tro }}}{f_{A}}$. 
	
	
	\section{Simulation experiment}
	The HPTFS was onboard CSS since October 2022. To verify that the dual-frequency transfer model is valid when the clock's long-term stability reaches the magnitude of $10^{-18}$, we conducted a simulation experiment to calculate the GP of GS. The duration of this experiment spanned one month,  from August 1 to 31, 2021, encompassing a total of 31 days. The orbits of CSS are computed using the two-line element sets (TLEs) data~\cite{riesing2015}. The ionospheric and tropospheric frequency shifts were assessed based on the models mentioned in section~\ref{method_2}. Additionally, we utilized the Earth Gravitational Model 2008 (EGM2008)~\cite{pavlis2012} to calculate the gravitational potential, GP, solid earth tide and other parameters essential for the experiment.  Furthermore, the clock error data of the CSS and GS also need to be simulated for this experiment.
	
	\subsection{Experimental procedure and setup}
	The Luojia time and frequency geodesic center (LJTF), located at Wuhan, China, was selected as the GS for the simulation experiment. As shown in table~\ref{tab_lj}, the coordinates of the LJTF are ($30^{\circ} 31^{\prime} 51.90274^{\prime \prime}\space \, \mathrm{N}$, $114^{\circ} 21^{\prime} 25.83516^{\prime \prime}\,  \mathrm{E}$, $25.728\, \mathrm {m}$). The gravitational potential and GP at this location are $62556081.21 \, \mathrm{m^2/s^2}$ and $62636467.54 \, \mathrm{m^2/s^2}$, respectively.
	
	\begin{table}[htb]
		\centering
		\begin{tabular}{|l|l|}
			\hline \text{ Parameters} & \text {Values} \\
			\text{ Latitude} & \text {$30^{\circ} 31^{\prime} 51.90274^{\prime \prime}\space \, \mathrm{N}$  } \\
			\text{ Longitude} & \text {$114^{\circ} 21^{\prime} 25.83516^{\prime \prime}\,  \mathrm{E}$ } \\
			\text{ Ellipsoid height (h)} & \text {$25.728\, \mathrm {m}$} \\
			\text{ Gravitational potential ($U$)} & \text {$62556081.21 \, \mathrm{m^2/s^2}$ } \\
			\text{ Gravity potential ($W$)} & \text {$62636467.54 \, \mathrm{m^2/s^2}$} \\
			\hline
		\end{tabular}
		\caption{The detailed information about the LJTF}
		\label{tab_lj}       
	\end{table}
	
	The observations in our simulation experiment are the frequency of the MWL carried by the CSS. These frequencies are denoted as $f_{A^{\prime}}$ and $f_{B}$ (where $f_{A^{\prime}}=f_{B}=30.4$ GHz), as described in Section~\ref{method}. By combining the emitted and received signals of $f_{A^{\prime}}/f_{B^{\prime}}$ and $f_{B}/f_{A}$ as illustrated in Figure~\ref{fig_MWL},  the difference of GP between CSS and LJTF can be measured.
	
	\begin{figure}[htb]
		\centering
		\includegraphics[width=0.85\textwidth]{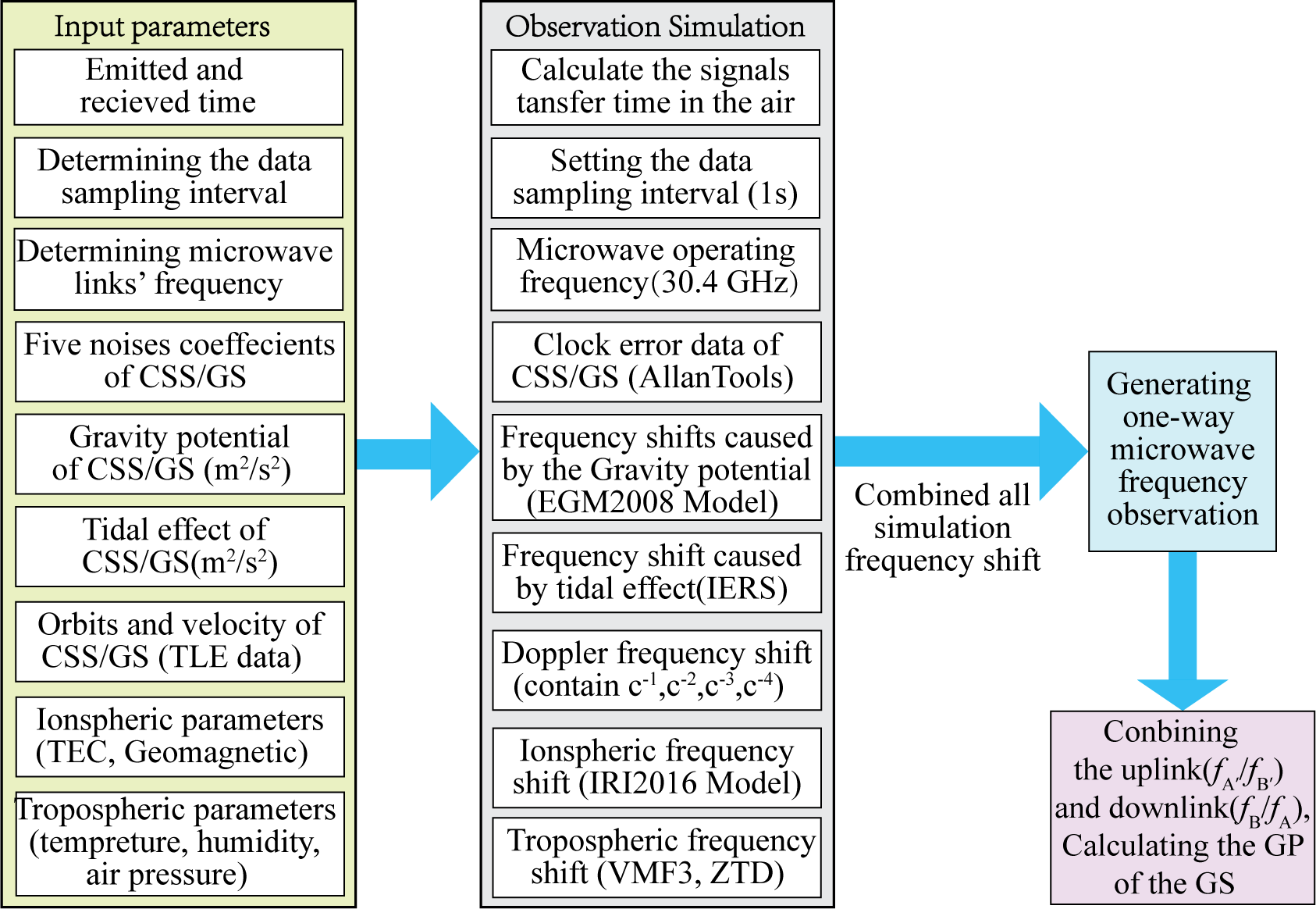}
		\caption{Simulation experiment parameter setting and flow chart. The first box on the left lists the various input parameters required for this simulation experiment. The second box in the middle outlines the process of generating the frequency shifts caused by these different parameters. Subsequently, all the parameters are integrated to produce the one-way MWL frequency observation. By combining the uplink and downlink signals, and dealing them with the proposed model, we can obtain the GP of the GS.}
		\label{fig_simulation} 
	\end{figure}
	
	As Figure~\ref{fig_simulation} shows,  the input parameters for our simulation experiments include the emitted and received signal times, the observation sampling interval, operating frequency point, atomic clock noises, GP and tidal effects of CSS and GS, ionospheric and tropospheric frequency shifts, as well as the orbits and velocity of CSS and GS. 
	We will now introduce the specific process of the simulation experiment in detail (see Figure~\ref{fig_simulation}). %
	The proper frequencies of uplink and downlink are set at 30.4 GHz with different polarization directions. This configuration can eliminate the tropospheric and ionospheric frequency shifts and extract the GRS. The sampling interval for the observation data is 1 second. Once the sampling interval and proper frequency are determined, we can simulate the frequency shifts by the theories introduced in Section~\ref{method}. First, we analyze the characteristics of the OAC and generate the clock error data of CSS and GS. The OACs' stabilities of the CSS and GS are $2\times10^{-15}$ and  $1\times10^{-15}$, respectively~\cite{Shen2023}. Next, we determine the positions and velocities of the CSS by the TLEs data. We apply the EGM2008 Model and international earth rotation service (IERS) parameters to calculate the GP and tidal effects for the CSS and GS,
	assessing their impact on frequency transfer ~\cite{pavlis2012,petit2010}. Following this, we estimate the Doppler frequency shifts, which include the terms of $c^{-1}$, $c^{-2}$, $c^{-3}$ and $c^{-4}$. We also consider the orbits of the CSS and GS while estimating the ionospheric and tropospheric frequency shifts. Since the troposphere and ionosphere affect the frequency of the MWL, their influences must be accounted for. In this experiment, the international reference ionosphere 2016 (IRI2016) model is used to calculate the input parameters when estimating the ionospheric frequency shifts~\cite{hoque2012, Zhangpf2023}. The tropospheric frequency shift is simulated by the zenith troposphere delay (ZTD) and the Vienna mapping function (VMF3)~\cite{Boisits2020, Landskron2018}.  Finally, by combining the uplink and downlink as described in equation (\ref{WB}), we obtain the GP of GS.
	
	\begin{figure}[!htbp]
		\centering
		\includegraphics{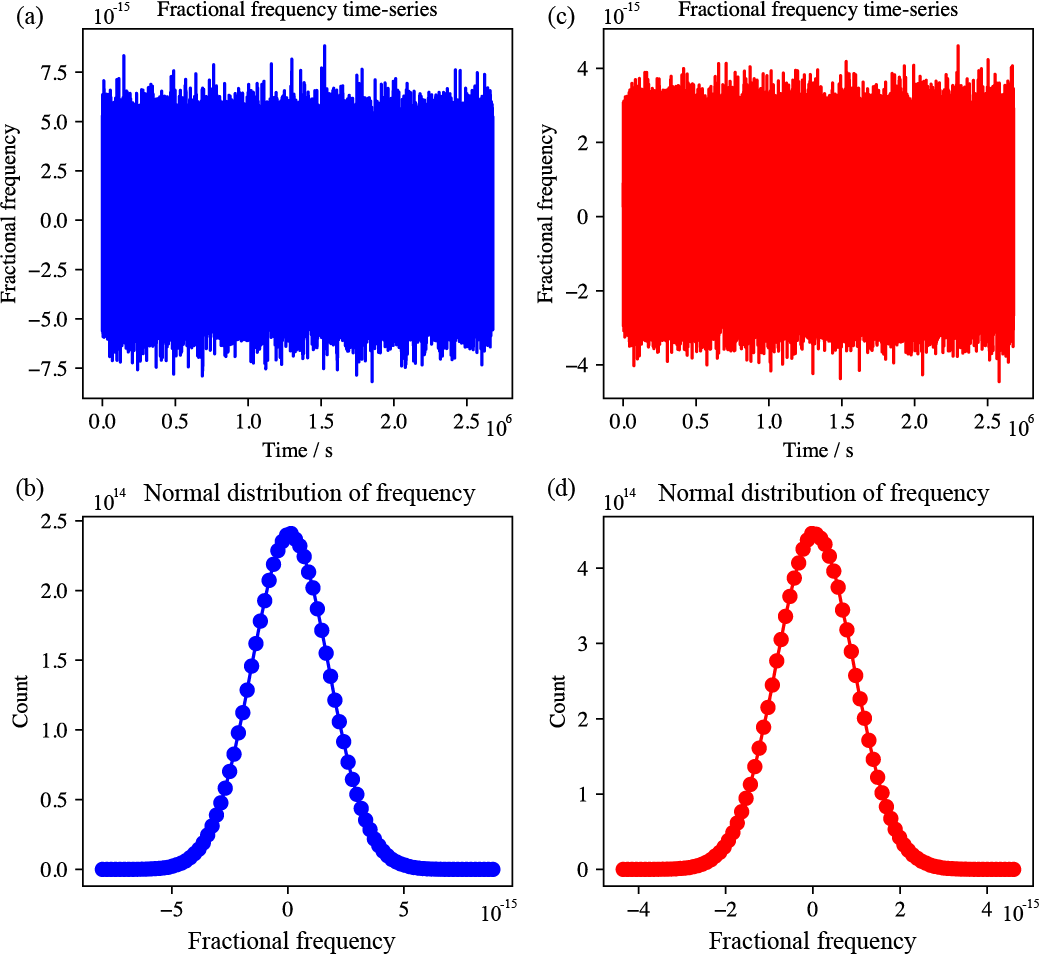}
		\caption{Simulation data of OACs generated by the clock error model. The blue and red lines represent the fractional frequencies of the OACs at CSS and GS, respectively. (a) and (c) display the fractional frequency fluctuations, (b) and (d) illustrate the normal distribution of fractional frequencies.}
		\label{fig_clocknoise}        
	\end{figure}
	
	From Section~\ref{method}, we know that $\nu_{S} / f_{S}$  and $f_{E}/\nu_{E}$ can be measured by the local clocks at stations S and E~\cite{blanchet2001}.
	To obtain the received and emitted frequency signals series, it is necessary to generate the clock noise data. Allan's research demonstrated that clock noises consist of five distinct types of random noise~\cite{allan1987,allan1991}. The clock frequency differences can be divided into deterministic and random components. The random component is represented as clock noise and different types of random noise are distinguished by the power spectral density (PSD) function of atomic clock frequency:
	\begin{equation}
		S_{y}(f)=\sum_{\alpha=-2}^{2} h_{\alpha} f^{\alpha} \label{psd}
	\end{equation} 
	where $S_{y}(f)$ is the PSD of the fractional frequency fluctuations, with units of $1/Hz$, here, $f$ is the Fourier frequency, $h$ is the intensity coefficient, and $\alpha$ is the exponent of the power-law noise process. The values of $\alpha=-2,-1,0,1,2$ correspond to different types of random noise:  random walk frequency modulation (RWFM), flicker frequency modulation (FFM), white frequency modulation (WFM), flicker phase modulation (FPM), and white phase modulation (WPM), respectively. The dominant noises in high-precision atomic clocks are WFM and RWFM~\cite{zucca2015}. In the experiment, the short-term stability of the OAC carried by the CSS is $2\times10^{-15}$(see Table~\ref{stability}), the short-term stability of the GS clock is $1\times10^{-15}$~\cite{Shen2023}. We simulated the clock fractional frequency noise series by a python library Allantools (\url{https://github.com/aewallin/allantools}).
	
	\begin{figure}[!htbp]
		\centering
		\includegraphics{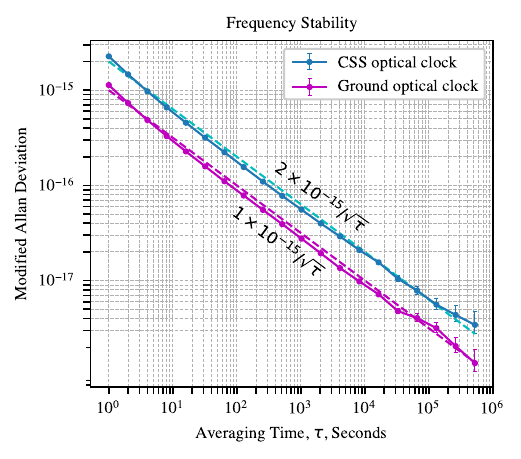}
		\caption{MDEV of the simulated clock errors. According to the design index, the stabilities of the CSS and GS optical clocks are approximate $2\times 10^{-15}/\sqrt{\tau}$ and  $1\times 10^{-15}/\sqrt{\tau}$.  The stability of CSS reaches $3\times10^{-17}$ after an accumulation time of 4096 seconds. The OACs' stabilities  of the CSS and GS reach $7.8\times10^{-18}$ and  $4.7\times 10^{-18}$ after $65500 $ seconds, respectively.}
		\label{fig_err}       
	\end{figure}
	
	Figure~\ref{fig_clocknoise} shows the fractional frequency signals generated by the simulation, which conform to the stochastic nature of clock noise. Figure \ref{fig_clocknoise}a and \ref{fig_clocknoise}b display the CSS optical clock noises, Figure \ref{fig_clocknoise}c and \ref{fig_clocknoise}d show the optical clock noises of the GS.  The clock noises of OACs in both the CSS and GS follow a normal distribution, with the CSS clock noise centered around $2\times10^{-15}$ and GS clock noises around $1\times10^{-15}$.
	
	In general, the modified Allan deviation (MDEV) is used as a time-domain characterization tool for assessing the frequency stability of atomic clocks~\cite{allan1987,allan1991}. To determine whether the generated clock noise data conforms to the design objectives, we use MDEV to evaluate the simulation data~\cite{allan1987,allan1991}.
	The MDEV of optical clock fractional frequency stabilities for the CSS and GS are shown in Figure~\ref{fig_err}. The fractional frequency stability of the CSS optical clock reaches $3\times10^{-17}@4000$ s which is similar to the designed value. When the evaluation time reaches $6.55\times10^4 $ seconds, the stabilities of CSS and GS optical clocks are $7.8\times10^{-18}$ and $4,7\times10^{-18}$, respectively. This indicates that our simulation data is in excellent agreement with the design values. We use these data to generate the emitted and received frequency signals.
	
	\begin{figure}[!htbp]
		\centering
		\includegraphics[width=0.75\textwidth]{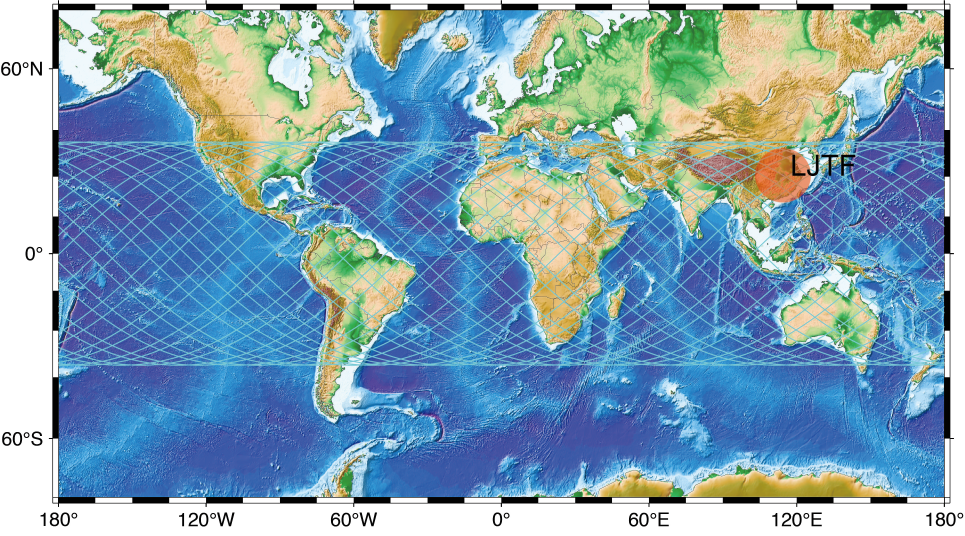}
		\caption{The trajectory of the CSS during two days of observation. The cyan links represent the CSS orbits with an inclination of $41.5^{\circ}$. The red circle is the visible range of the LJTF when the cutoff elevation angle is set to $15^{\circ}$.}
		\label{fig_orbits} 
	\end{figure}
	
	Other frequency shifts are influenced by the position and velocity of the CSS and GS.
	The CSS passes through the GS about four or five times a day. TLEs orbital data can be obtained from the website CelesTrak (\url{http://www.celestrak.com/NORAD/elements/stations.txt}) with the satellite name TianHe. The position, velocity, and acceleration of the CSS are calculated by the TLEs and used as the real orbital data. 
	Figure~\ref{fig_orbits} illustrates two days of tracking orbits for the CSS (see the cyan lines). In the simulation experiments, the cutoff elevation angles were set to $5^{\circ}$, $10^{\circ}$ and $15^{\circ}$, respectively. The range of the subsatellite point is represented by the red circle in Figure~\ref{fig_orbits} when the cutoff elevation angle is $15^{\circ}$. For the different cutoff elevation angles from $5^{\circ}$ to $15^{\circ}$, the duration during which the CSS passes directly overhead (zenith)  ranges from 500 seconds to 300 seconds.
	According to the design requirements for the CSS, the orbit determination accuracy is $\pm 0.1 \mathrm{~m}$  and velocity measurement accuracy reaches $\pm 1.0 \times 10^{-3} \mathrm{~m}$ after processing~\cite{wang2021}. These two measurement errors are considered white noise and are added to the true values in the simulation experiments.
	The theories of Section~\ref{method_2} and ~\ref{method_3} are created in the geocentric inertial coordinate frame, which is an Earth-centered inertial (ECI) frame. To facilitate this, a coordinate transformation from the WGS84 (World Geodetic System 1984) to the ECI is performed using the Earth orientation parameter (EOP) data download from IERS website (\href{https://datacenter.iers.org/data/latestVersion/223_EOP_C04_14.62-NOW.IAU1980223.txt}{https://datacenter.iers.org/data/latestVersion/})~\cite{petit2010}.
	
\begin{figure*}[htb]
		\centering
		\includegraphics[width=1\textwidth]{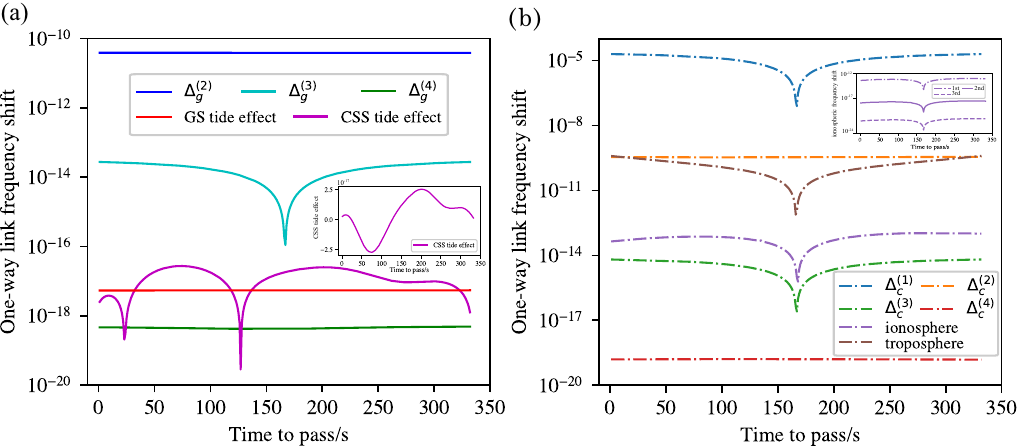}
		\caption{All frequency shifts caused by different effects in one-way frequency transfer. (a) shows the frequency shift caused by the GP and tidal effects of CSS and GS, the subgraph illustrates the detail of the CSS's tidal effect. (b) depicts the Doppler frequency shifts caused by the position, velocity, troposphere and ionosphere, the subgraph shows the first-order, second-order and third-order ionospheric frequency shifts.}
		\label{fig_shift}       
	\end{figure*}

	After obtaining the position and velocity of the CSS and GS, we can calculate the GP and tidal effects through the EGM2008 model. According to the equation~\ref{dop_g}, the frequency shifts caused by the GP and tidal influences can be simulated.  The gravity and gravitational potential of the CSS and LJTF 
	derived from the EGM2008 model~\cite{pavlis2012}. For these calculations, we utilize the WGS84 reference ellipsoid. The tidal potentials for both the CSS and GS are determined by the orbital positions of the sun and moon, using only the second-order Legendre polynomial and corrected for the solid earth tide by using the love numbers~\cite{munk1966,cartwright1977}. Since the indirect tide is smaller than the residual errors, it is not considered in the simulation experiments.  {Figure~\ref{fig_shift}a} displays the absolute values of frequency shifts affected by the GP and tidal effects. $\Delta_{g}^{(2)}$, $\Delta_{g}^{(3)}$ and $\Delta_{g}^{(4)}$ are the frequency shifts caused by the second-order, third-order and fourth-order of the GP, with magnitudes reaching about $10^{-11}$, $10^{-14}$ and $10^{-19}$, respectively. The tidal effects for CSS and GS lead to the frequency shifts on the magnitude of $10^{-17}$. Simultaneously, the frequency shifts caused by the Doppler effect can also be simulated through equation~(\ref{dop_c}). {As shown in Figure~\ref{fig_shift}b},  the Doppler frequency shifts $\Delta_{c}^{(1)}$, $\Delta_{c}^{(2)}$, $\Delta_{c}^{(3)}$ and $\Delta_{c}^{(4)}$ reach $10^{-5}$, $10^{-10}$, $10^{-15}$ and $10^{-19}$, respectively.
	
	When the MWL passes through the atmosphere, the resulting frequency shifts are influenced by various factors, including atmospheric effects (troposphere and ionosphere), Earth's solid tide, the special relativistic Doppler effect, and contributions from the gravitational field. Frequency shifts in a vacuum can be calculated by using equations~(\ref{1}) to (\ref{lb}). However, since the CSS orbit altitude is about $350\sim450$ km, and the MWL emitted or received at the CSS is primarily affected by the troposphere and ionosphere below this altitude. To estimate the ionospheric frequency shifts, Total Electron Content (TEC) data are obtained from the Internation Reference Ionosphere 2016 (IRI2016) model~\cite{liu2019}. The integration is carried out from the bottom of the ionosphere up to the CSS orbital altitude~\cite{Zhangpf2023}. For second-order ionospheric frequency shifts, the geomagnetic field is calculated by the International Geomagnetic Reference Field (IGRF)~\cite{alken2021}. By combining the TEC with the ionospheric mapping function, the slant TEC can be obtained, and the total ionospheric frequency shifts can be simulated by using equation~(\ref{dfion}). The  magnitude of ionospheric frequency shifts range from about  $10^{-13} \sim 10^{-15}$ (as shown in Figure~\ref{fig_shift}b). 
	The tropospheric frequency shifts can be evaluated by using equation~(\ref{dftro}),   for which parameters such as temperature, dry pressure, and wet pressure are required. These parameters can be downloaded from the website of  Crustal Dynamics Data Information System(CDDIS, \url{https://cddis.nasa.gov/archive/gnss/data/daily/2021/212/21m/}). In practice, we use the VMF3 and ZTD to estimate the slant tropospheric delay~\cite{Boisits2020, Landskron2018}, and after that,  convert this into the frequency shifts. The magnitude of the tropospheric frequency shifts ranges from approximately $10^{-10} \sim 10^{-13}$(see Figure~\ref{fig_shift}b).
	
	
	By simulating all these frequency shifts and combining them with the up and down MWLs at proper frequency $30.4 \, \mathrm{GHz}$, we can obtain the observation data (See Figure~\ref{fig_simulation}). We generate all 31 days of data for the simulation experiment. The GP of the LJTF can be determined by the dual-frequency transfer model (see equation \ref{WB}) proposed in this study.
	
	\subsection{Results}
	
	We calculate 31 days of data from August 1 to 31, 2021. For the sake of description, each time the CSS passes over the GS, it is defined as one observation, and we can obtain a series of records with 1-second sampling for each observation. Figure~\ref{fig_shift} shows all the frequency shifts for one observation. Due to the relative motion between the CSS and LJTF, the up and down MWLs do not coincide exactly. When we use the dual-frequency combination method to deal with the observations, there are some residual {errors, which are corrected using relevant models.}
	
	\begin{table}[htb]
		\centering
		\begin{tabular}{|l|l|c|}
			\hline \text{ Error Type } & \text { Magnitude of errors } & \text { Residual errors } \\
			\hline 	
			\text { Doppler frequency shift } & $10^{-5} \sim 10^{-6}$ & $<1.7 \times 10^{-17}$ \\
			\text { Shapiro effect ($\Delta_{c}^{(3)}$) } & $\sim10^{-15} $ & $ \sim 10^{-22} $\\
			\text { Ionospheric frequency shift } & $10^{-13} \sim 10^{-15}$ & $< 1.5\times 10^{-19} $\\
			\text { Tropospheric frequency shift } & $10^{-10} \sim 10^{-13}$ & $<2.1 \times 10^{-17}$ \\
			\text { Total Tidal potential} &$< 3.5 \times 10^{-17}$ &$<1.0 \times 10^{-18}$ \\
			\text { Total frequency shift } &$10^{-5} \sim 10^{-6}$ &$<4.3\times 10^{-17}$ \\
			\hline
		\end{tabular}
		\caption{Different types of errors and  dual-frequency transfer correction residual errors }
		\label{tab_err}       
	\end{table}
	
	Table~\ref{tab_err} shows the magnitudes of Doppler frequency shift, ionospheric and tropospheric frequency shifts, which are approximately $10^{-5} \sim 10^{-6}$, $10^{-13} \sim 10^{-15}$, $10^{-10} \sim 10^{-13}$, respectively. The Shapiro effect frequency shift is about $ 10^{-15}$. After processing the observation data with the dual-frequency combination model, it is essential to eliminate errors with the relevant models. The residual errors after the tropospheric and ionospheric model corrections reach 5\%~\cite{hoque2012} and 10\%~\cite{saastamoinen1972, Landskron2018} respectively. The Doppler effect can be eliminated by using equation~(\ref{dop_c}). The tidal potential can be corrected by the HW95 catalog \cite{hartmann1995}. When the value of Legendre function degree $\ell$ is greater than 3, the maximum tidal potentials of the Moon and Sun are $7.88 \times 10^{-2}\, \mathrm{m^2/s^2}$ and $6.80 \times 10^{-5}\, \mathrm{m^2/s^2}$, respectively~\cite{hartmann1995}. In the simulation experiment, the residual frequency shift caused by the tidal potential is less than $1\times10^{-18}$. The residual errors of various frequency shifts are detailed in Table~\ref{tab_err}. The sum of these frequency shifts is less than the magnitude of $10^{-15}$, that is to say, after correcting with the observations, the main errors come from the clock errors. Hence, the stability of the OAC determines the GP measurement accuracy. 
	
	\begin{figure}[htb]
		\centering
		\includegraphics[width=.55 \textwidth]{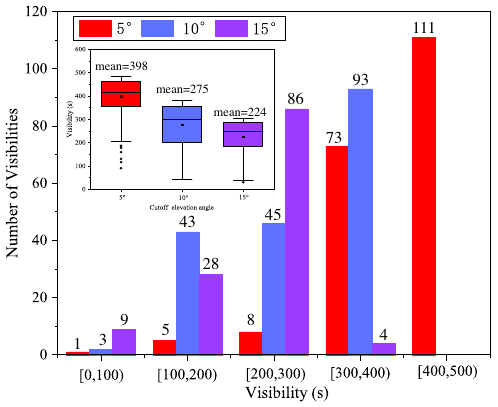}
		\caption{The observation numbers and duration of each observation for different cutoff elevation angles. The colors in red, blue, and purple are the number of observations when the cutoff elevation angles are $5^{\circ}$, $10^{\circ}$ and $15^{\circ}$ respectively. When the cutoff elevation angle is $5^{\circ}$, most observation time is in the range of  $(300\sim500)$ s. If the cutoff elevation angle is $10^{\circ}$, the observation duration is distributed between 200 s and 400 s. /$(100\sim300)$ s. When the cutoff elevation angle is set as $15^{\circ}$, most of the observation duration is less than 300 s.  For all these three setup cutoff elevation angles, only a few observation times are less than 100 s.}
		\label{fig_obs}       
	\end{figure}
	
	As the CSS flies along its orbit, when we set the different cutoff elevation angles, the length of observation time ranges from dozens of seconds to approximately $500 \,\mathrm{s}$. Figure~\ref{fig_obs} shows when the cutoff elevation angle is set to $15^{\circ}$, there are 127 observations over 31 days, with most observation durations are distributed around $200 \sim 300 \, s$.  
	From the characteristics of ADEV/MDEV, it is known that the frequency stability of atomic clocks improves with increased accumulation time until FFM and RWFM become the dominant noise factors~\cite{allan1987,allan1991}. Therefore, extending the observation time can effectively enhance measurement accuracy. As observation time lengthens with a decrease in the cutoff elevation angle, we can increase the observation duration by lowering the cutoff elevation angle. The distributions of observation times for different cutoff elevation angles are shown in Figure~\ref{fig_obs}. The number of observations is 198, 184, and 127 when the cutoff elevation angles are $5^{\circ}$, $10^{\circ}$ and $15^{\circ}$ respectively. 
	
	\begin{figure}[htb]
		\centering
		\includegraphics[width=1 \textwidth]{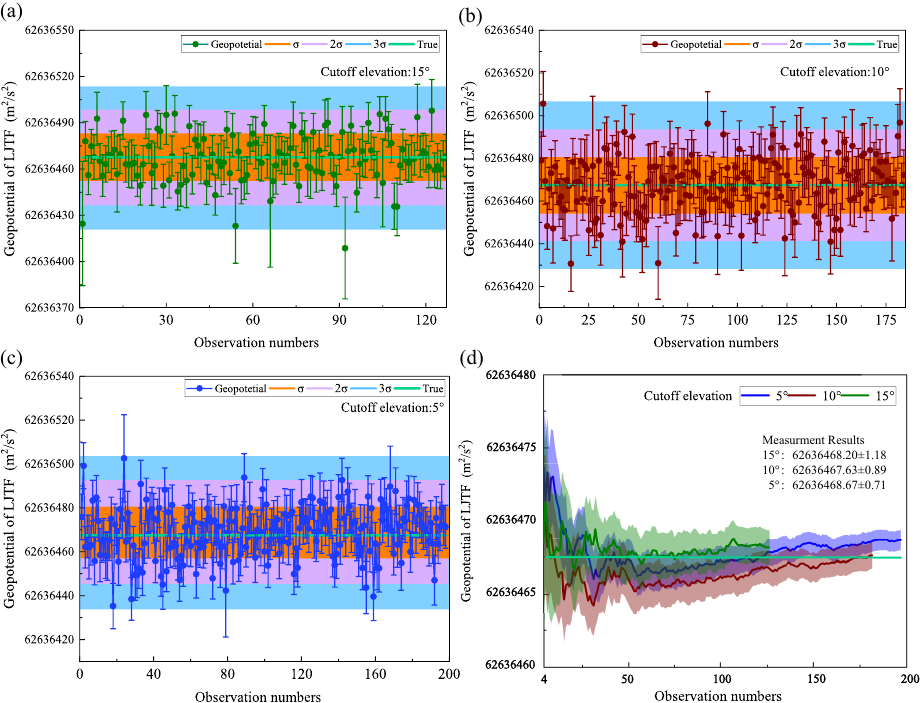}
		\caption{The results of GP calculated using dual-frequency observation data at different cutoff elevation angles. Subgraphs (a), (b), and (c) show the results of GP calculated by the observations at the cutoff elevation angle of  $15^{\circ}$,  $10^{\circ}$. and  $5^{\circ}$, respectively. Subgraph (d) presents the statistical results of GP calculated at different cutoff elevation angles.}
		\label{fig_result}       
	\end{figure}
	
	In Figure~\ref{fig_obs}, the colors in red, blue, and purple represent the number of observations for the cutoff elevation angles of $5^{\circ}$, $10^{\circ}$ and $15^{\circ}$ respectively.  From Figure~\ref{fig_obs}, it can be observed that at a cutoff elevation angle of $5^{\circ}$, there are 73 observation durations in the range of 300 s to 400 s, as well as 111 observation durations exceeding 400 s, the total numbers of these two observation values accounts for 92.9\% of all observations. When the cutoff elevation angle is $10^{\circ}$, most of the observation durations fall within in $200\sim300$ s range, representing 75.0\% of total observations. When the cutoff elevation angle is $15^{\circ}$, only four observation data have durations exceeding 300 s. By analyzing these observation datasets and calculating the average observation durations for different cutoff elevation angles, we found that when the elevation angles are $5^{\circ}$, $10^{\circ}$, and $15^{\circ}$, the average observation durations are 398 s, 275 s, and 224 s, respectively. 
	
	
	
	We calculate the GP of the LJTF by the dual-frequency transfer method for different cutoff elevation angles. Figure~\ref{fig_result} displays the results of GP  obtained from dual-frequency observation data at different cutoff elevation angles. It uses the green, brown, and blue lines to present the results calculated by different observations measured in $15^{\circ}$,  $10^{\circ}$ and  $5^{\circ}$. The range of $\sigma$, $2\sigma$, and $3\sigma$ are represented by the claybank, light-pink, and light-blue shaded areas, respectively. Previous studies have shown that the $\sigma$ criterion is an effective method for identifying and eliminating gross error in measurements~\cite{su2009}. When the observation durations are sufficiently long and all measurement accuracies are equal, random errors typically fall within the ranges of $[-3\sigma,3\sigma]$ and $[-2\sigma,2\sigma]$ for $99.73\%$ and $95.44\%$ of all random errors, respectively.
	Through analysis of the calculation results, the  $\sigma$ of GP for $15^{\circ}$,  $10^{\circ}$ and  $5^{\circ}$ are $15.57\mathrm{~m^2/s^2}$, $13.19\mathrm{~m^2/s^2}$ and $11.77\mathrm{~m^2/s^2}$, respectively. The values of $\sigma$  decrease with the cutoff elevation angles decrease. From Figure \ref{fig_result}a, \ref{fig_result}b, and \ref{fig_result}c, it is evident that, except for the case with a cutoff elevation angle of $15^{\circ}$, all other observations fall within the range of $3\sigma$. Due to the different precision of each observation, we use the weighted average formula to calculate the final GP and its standard deviation (STD). Figure \ref{fig_result}d shows the final calculation results. It can be seen from the figure that when the cutoff angles are $15^{\circ}$, $10^{\circ}$ and  $5^{\circ}$, the measured GP of LJTF are $(62636468.20\pm1.18 )\, \mathrm{m^2/s^2}$, $(62636467.63\pm0.89)\, \mathrm{m^2/s^2}$, and $(62636468.67\pm0.71)\, \mathrm{m^2/s^2}$, respectively. By analyzing the observation numbers and durations (as shown in Figure~\ref{fig_obs} and \ref{fig_result}),  we find that longer observation durations lead to higher availability and more reliable measurement results.

	\begin{table}[htb]
		\centering
		\begin{tabular}{|c|c|c|c|}
			\hline  \text {Cutoff elevation angle} & $\mathrm{GP}\left(\mathrm{m}^2 / \mathrm{s}^2\right)$ & Bias $\left(\mathrm{m}^2 / \mathrm{s}^2\right)$ & STD $\left(\mathrm{m}^2 / \mathrm{s}^2\right)$ \\
			\hline $5^{\circ}$ & 62636468.67& 1.13& $\pm 0.71$ \\
			$10^{\circ}$ & 62636467.63& 0.09 & $\pm 0.89$ \\
			$15^{\circ}$ &  62636468.20& 0.66 & $\pm 1.18$ \\
			\hline
		\end{tabular}
		\caption{GP of LJTF calculated by the simulation experiments.}
		\label{tab_result}       
	\end{table}
	
	Table~\ref{tab_result} presents the results of the simulation experiment for different cutoff elevation angles. 
	The biases of the GP, calculated by the observations at cutoff elevation angles of $5^{\circ}$,  $10^{\circ}$. and  $15^{\circ}$ are $1.13 \, \mathrm{m^2/s^2}$, $0.09 \, \mathrm{m^2/s^2}$ and $0.66 \, \mathrm{m^2/s^2}$, respectively.  
	The STD  for cutoff elevation angles $5^{\circ}$, $10^{\circ}$ and $15^{\circ}$ are $\pm0.71 \, \mathrm{m^2/s^2}$, $\pm0.89, \mathrm{m^2/s^2}$ and $\pm1.18 \, \mathrm{m^2/s^2}$, respectively. Notably, the STD decreases as the cutoff elevation angle decreases.

	\section{Conclusion}
	
	In this study, we present a one-way frequency transfer model based on frequency transfer up to the order $c^{-4}$ in the free space, with an accuracy reaching $10^{-19}$. The model takes into account the effects of the troposphere, ionosphere, and Earth solid tide on the MWL's frequency. To eliminate these influences, we build a dual-frequency transfer model based on the characteristics of CSS microwave links to determine the GP of the GS. The highlight of this model is that the up and down MWL signals have the same frequency but with different polarization directions. This design significantly reduces the first-order Doppler effect, as well as the impacts of the troposphere and ionosphere, especially the second-order terms. Since the up and down MWLs are transferred almost simultaneously, the dual-frequency model can largely eliminate the Doppler frequency shifts, tropospheric and ionospheric frequency shifts, and other frequency shifts caused by the MWL propagation. After processing the frequency data with the dual-frequency combination model, the main source of error in the observations comes from the OAC. The accuracy of the OAC directly determines the measurement accuracy of the GP.
	
	We selected the LJTF as the GS and conducted a simulation experiment to verify the validity of the model. The experiment used 31 days of data to analyze the influence of different cutoff elevation angles on the observations. In the experiment, the frequency can be measured by the FOFC and doesn't need to make the phase ambiguities fixed. We analyzed how different cutoff elevation angles affect both the measurement duration and accuracy.
	According to the statistics, when the cutoff elevation angles are $5^{\circ}$,  $10^{\circ}$, and $15^{\circ}$, the corresponding average observation durations are 398 s, 275 s, and 224 s, respectively. The observation duration decreases as the cutoff elevation angle increases. We calculated the GP of the LJTF and assessed the accuracy of the results for different cutoff elevation angles. If the cutoff elevation angle is $15^{\circ}$, the GP of the LJTF is  $(62636468.20\pm1.18)\,\mathrm{m^2/s^2}$, with a bias of $0.66\,\mathrm{m^2/s^2}$. When the cutoff elevation angles are set to $5^{\circ}$ and $10^{\circ}$, the observation duration increases to about $300\sim500$ s, resulting in more available data. The results of GP for $5^{\circ}$ and  $10^{\circ}$ cutoff elevation angles were $(62636468.67\pm0.71)\,\mathrm{m^2/s^2}$ and $(62636467.63\pm0.89)\,\mathrm{m^2/s^2}$,  with biases of $1.13\,\mathrm{m^2/s^2}$ and $0.09\,\mathrm{m^2/s^2}$, respectively. The measurement accuracy of the GP improves as the cutoff elevation angle decreases.

	This study demonstrates that the proposed model allows for the measurement of the GP with centimeter-level accuracy. It is the first detailed study of the dual-frequency transfer model, supported by a simulation experiment. This work introduces a novel approach for future frequency transfer. With the development of the OAC, the CSS will be equipped with a Yb OAC whose long-term stability reaches $10^{-19}$~\cite{lv2021}, and the proposed model could be applied to GP determination at centimeter level. 
	
	\acknowledgments
	
	This study is supported by the National Natural Science Foundation of China (NSFC)(Grant Nos. 42388102, 42030105,42274011). Funded by State Key Laboratory of Geo-Information Engineering and Key Laboratory of Surveying and Mapping Science and Geospatial Information Technology of MNR, CASM: 2024-01-01, the China Postdoctoral Science Foundation(Certificate Number: 2024M752480 ).
	
	\paragraph{Open Access.} This article is distributed under the terms of the Creative Commons
	Attribution License (CC-BY 4.0), which permits any use, distribution and reproduction in
	any medium provided the original author(s) and source are credited.
	
	
	\bibliographystyle{JHEP}
	\bibliography{ref.bib}
	
\end{document}